\documentclass[aps,prl,twocolumn,superscriptaddress,floatfix,nofootinbib,showpacs,longbibliography]{revtex4-1}

\usepackage[utf8]{inputenc}  
\usepackage[T1]{fontenc}     %Output what you want e.g., é, ł, a, ü
\usepackage[british]{babel}  %Do hyphenation according to british english
\usepackage[sc,osf]{mathpazo}
\usepackage[scaled=0.86]{berasans}  % URL font that go well wtih palatino
\usepackage[colorlinks=true, citecolor=blue, urlcolor=blue]{hyperref}  
\usepackage{graphicx} % Package to insert exteral figures
\usepackage[babel]{microtype}  %Improves text justification
\usepackage{amsmath,amssymb,amsthm,bm,amsfonts,mathrsfs,bbm} %Usefull math packages

\usepackage{xspace}  %Useful to add space in macros
\usepackage{pgf,tikz}
\usepackage{xcolor}
\usepackage{multirow}
\usepackage{array}
\usepackage{bigstrut}
\usepackage{braket}
\usepackage{color}
\usepackage{natbib}
\usepackage{multirow}
\usepackage{mathtools}
\usepackage{float}
\usepackage[caption = false]{subfig}
\usepackage{xcolor,colortbl}
\usepackage{color}
\newcommand{\Tr}{\operatorname{Tr}}

\newcommand{\be}{\begin{equation}}
\newcommand{\ee}{\end{equation}}
\newcommand{\ba}{\begin{eqnarray}}
\newcommand{\ea}{\end{eqnarray}}

\newcommand{\tr}{\operatorname{Tr}}

\def\>{\rangle}
\def\<{\langle}

\usepackage{centernot}
\usepackage{subfig}

\begin{document}

\title{Better transmission with lower capacity: lossy compression over quantum channels}

\author{Sristy Agrawal}
\affiliation{Department of Physics and Center for Theory of Quantum Matter, University of Colorado, Boulder, Colorado 80309, USA}
\affiliation{JILA, University of Colorado/NIST, 440 UCB, Boulder, CO 80309, USA}

\affiliation{National Institute of Standards and Technology, Boulder, Colorado 80305, USA}

\author{Rajashik Tarafder}
\affiliation{The Division of Physics, Mathematics and Astronomy, California Institute of Technology, Pasadena CA 91125, USA}
\affiliation{Institute of Quantum Information and Matter, California Institute of Technology, Pasadena CA 91125, USA}

\author{Graeme Smith}
\affiliation{Department of Physics and Center for Theory of Quantum Matter, University of Colorado, Boulder, Colorado 80309, USA}
\affiliation{JILA, University of Colorado/NIST, 440 UCB, Boulder, CO 80309, USA}

\author{Arup Roy}
\affiliation{Department of Physics, A B N Seal College Cooch Behar, West Bengal 736101, India.}

\author{Manik Banik}
%\email{manik11ju@gmail.com}    
\affiliation{School of Physics, IISER Thiruvananthapuram, Vithura, Kerala 695551, India.}	
	
\begin{abstract}
Shannon's channel coding theorem describes the maximum possible rate of reliable information transfer through a classical noisy communication channel. It, together with the source coding theorem, characterizes lossless channel communication in the classical regime. Lossy compression scenarios require the additional description provided by rate-distortion theory, which characterizes the tradeoff between compression rate and the distortion of the compressed signal.  Even in this context, the capacity characterizes the usefulness of a channel---a channel with more capacity will always outperform a channel with less capacity. We show that this is no longer true when sending classical information over a quantum channel.  In particular, we find a pair of quantum channels where the channel with the lower capacity causes less distortion than the higher capacity channel when both are used at a fixed rate. 
\end{abstract}

%\pacs{03.65.Ta,03.65.Ud, 03.67.Dd, 03.67.Hk}
%\keywords{\textclor{red}{Need to be added}}

% 03.65.Ta	Foundations of quantum mechanics;
% 03.67.Dd	Quantum cryptography and communication security
% 03.67.Hk	Quantum communication
% 03.65.Ud	Entanglement and quantum nonlocality

\maketitle	
%\section{Introduction}
{\it Introduction.--} The basic purpose of any communication protocol is transfer of information from one space-time point to another while safeguarding the information in-transit against an assumed error model, {\it i.e.}, a channel. A classical channel is modeled by a stochastic map from input random variable $X$ to output random variable $Y$; and, its capacity is given by the mutual information - which is an entropic quantity - optimized over the probability distributions of the input variables \cite{Shannon48}. In a quantum scenario, a channel is described by a completely positive trace preserving (CPTP) super-operator that maps operators on some input Hilbert space to the operators on some output Hilbert space. Depending on the information-theoretic task, different quantities of interest can be defined to characterize the utility of a quantum channel. For instance, quantum capacity of a quantum channel denotes the highest rate of transmitting quantum information \cite{LSD1,LSD2,LSD3}, and its private capacity provides the rate of transferring classical information privately \cite{LSD3}, over many uses of a noisy quantum channel. Some of these communication methods have no classical analogues and make quantum information theory much richer than its classical counterpart.  

On the other hand, similar to classical channels, a quantum channel can also be used to send classical information; and the relevant quantity of interest, in that case, is called the classical capacity. Interestingly, while sending classical information, a quantum channel can exhibit peculiar features that are not possible in a classical channel. For instance, the additivity of classical mutual information makes repeated use of a single classical channel and parallel use of many copies of that channel identical \cite{Cover}. But, in the quantum scenario, the quantum entanglement leads to the super-additivity of a quantum channel's classical capacity. Depending on whether entanglement is allowed during encoding or/and decoding  (while many copies of the channel are used in parallel) classical capacity of a quantum channel can be defined in various ways - namely, Shannon capacity $(\mathrm{C}_{Shan})$, Holevo capacity $(\mathrm{C}_{Hol})$ and Ultimate capacity $(\mathrm{C}_{ult})$ \cite{Bennett98} (see also \cite{King01}). While $\mathrm{C}_{Shan}$ quantifies the classical capacity of a quantum channel with product encoding and product decoding, $\mathrm{C}_{Hol}$ captures the same but with entangled measurement allowed during the decoding process \cite{Schumacher97,Holevo98}. Importantly, there exist channels for which $\mathrm{C}_{Hol}>\mathrm{C}_{Shan}$  \cite{Fuchs17,Schumacher01,King02,Hayashi05}, indicating the signature of supper-additivity while transferring classical information through a quantum channel. It is worth mentioning that both quantum capacity ($\mathrm{Q}$) and private capacity ($\mathrm{P}$) exhibit the super-additivity phenomenon \cite{Smith08,Oppenheim08,Smith11,Li09,Smith09}; and an unbounded number of channel uses may be considered to obtain the capacity $\mathrm{Q}$ \cite{Cubitt15}. In-fact, entangled input states are not necessary for $\mathrm{Q}$ to be super-additive \cite{Leditzky18,Yu20}. But, if entanglement is allowed at input, then even the Holevo capacity of a quantum channel can exhibit super-additivity feature \cite{Hastings09}. The capacity is now defined as  $\mathrm{C}_{ult}$ which captures classical capacity of a quantum channel while entanglement is allowed both for encoding and decoding (the authors in \cite{King01} have referred to this as {\it entangled signals/entangled measurements Capacity}).   

This super-additivity of the rate of information transfer makes the study of quantum Shannon theory much more exciting than its classical counterpart. However, the extreme difficulty of finding a single letter expression for $\mathrm{Q}$, $\mathrm{P}$, and $\mathrm{C}_{ult}$ makes  the study of quantum Shannon theory particularly challenging. Nonetheless, the quantity $\mathrm{C}_{ult}$ seems to capture the usefulness of a quantum channel for classical information transfer when no pre-shared correlation is allowed between the sender and the receiver \cite{Bennett92,Bennett99,Bennett02}. Given two noisy quantum channels $\mathcal{N}_1$ and $\mathcal{N}_2$ let the ordering relation $\mathcal{N}_1\preceq\mathcal{N}_2$ indicates superiority of the channel $\mathcal{N}_2$ over $\mathcal{N}_1$, {\it i.e.} the channel $\mathcal{N}_2$ is at least as good as the $\mathcal{N}_1$ for any classical information processing task. It seems natural to assume that $\mathcal{N}_1\preceq\mathcal{N}_2$ whenever $\mathrm{C}_{ult}(\mathcal{N}_1)\le\mathrm{C}_{ult}(\mathcal{N}_2)$. Quite surprisingly, in this paper, we establish that this is not always the case. We find that the quantum channel $\mathcal{N}_1$ may be preferable over the channel $\mathcal{N}_2$ in certain scenarios even though the former has a smaller classical capacity than the latter one.  Particularly, we show that there exist operational tasks involving two separated players achieving higher collaborative payoff $\theta_{\$}$ with a quantum channel of less classical capacity, {\it i.e.} $\theta_{\$}(\mathcal{N}_1)>\theta_{\$}(\mathcal{N}_2)$ while $\mathrm{C}_{\pounds}(\mathcal{N}_1)<\mathrm{C}_{\pounds}(\mathcal{N}_2)$ for $\pounds\in\{Shan,~Hol,~ult\}$. This intriguing feature is demonstrated with explicit examples of qubit channels. 

Furthermore, using rate distortion theory, we establish that the origin of this peculiarity is truly quantum; as such a phenomenon is not possible with classical channels.  Our result thus adds a new twist in the study of quantum channel capacities by making their utility in classical information transfer context dependent.

%\section{Preliminaries}
{\it Preliminaries.--} A quantum discrete memoryless channel (QDMC) is described by a CPTP map $\mathcal{N}$ acting on a quantum information carrier associated with some Hilbert space $\mathcal{H}_{in}$. The quantum state $\rho\in\mathrm{D}(\mathcal{H}_{in})$ evolves under the action of $\mathcal{N}$; where $\mathrm{D}(\mathcal{H}_{in})$ is the set of all density operators on $\mathcal{H}_{in}$. This evolution can most generally be described as $\rho\mapsto\mathcal{N}(\rho)=\Tr_E[U(\rho\otimes\sigma)U^\dagger]$; where $\sigma$ is some fixed state of the environment, $U$ is some unitary operator on the joint Hilbert space of carrier and environment, and $\Tr_E$ denotes a partial trace over the environmental degrees of freedom \cite{Chuang00}. It is also convenient to express a quantum channel in Kraus form $\mathcal{N}(\rho) = \sum_j A_j\rho A_j^\dagger$; where the Kraus operators $A_j: \mathcal{H}_{in}\rightarrow \mathcal{H}_{out}$ are linear operators satisfying $\sum_jA_j^\dagger A_j=\mathbf{I}_{in}$, $\mathcal{H}_{out}$ is the output Hilbert space, and $\mathbf{I}_{in}$ is the identity operator on $\mathcal{H}_{in}$ \cite{Kraus}. For a memoryless channel the evolution for arbitrary states $\eta\in\mathrm{D}(\mathcal{H}_{in}^{\otimes n})$ is given by $\mathcal{N}^{\otimes n}(\eta)=\sum_{j_1\cdots j_n}(A_{j_1}\otimes\cdots\otimes A_{j_n})\eta(A_{j_1}^\dagger\otimes\cdots\otimes A_{j_n}^\dagger)$.

While transmitting classical information through a QDMC, the sender (let's call them Alice) encodes the messages $x\in X$ into some pure state $\eta_x\in\mathrm{D}(\mathcal{H}_{in}^{\otimes n})$ which may be entangled across the $n$ subsystems. The intended receiver (let's call them Bob), receives the evolved state $\mathcal{N}^{\otimes n}(\eta_x)$ and performs some measurement $\{E_y~|~E_y\ge0~\&~\sum_{y}E_y=\mathbf{I}_{out}^{\otimes n}\}$ (a positive operator-valued measure - POVM) to guess the quantum state $\eta_x$ and decodes $x$. Similar to the encoding, entanglement is also allowed at the decoding step in general, {\it i.e.} the POVM elements may be a collective quantum measurement over all the qubits and need not factorize into measurements on the individual carriers. As noted by Bennett and Shor \cite{Bennett98}, depending on how entanglement is used during encoding and decoding, four types of channel capacities can be defined -- $\mathrm{C}_{\alpha\beta}$ with $\alpha,\beta\in\{P\equiv\mbox{Product},E\equiv\mbox{Entangled}\}$; where $\alpha$ and $\beta$ respectively denote what kind of encoding and decoding are used. In general, we have $\mathrm{C}_{PP}=\mathrm{C}_{EP}\le\mathrm{C}_{PE}\le\mathrm{C}_{EE}$. The first equality is  established in \cite{King01}. The expression for the product Capacity $\mathrm{C}_{PP}$, also known as Shannon Capacity $\mathrm{C}_{Shan}$, is given by     
\small
\begin{equation}
\mathrm{C}_{Shan}(\mathcal{N})=\sup_{\mathcal{E}}\sup_{\{E_b\}}\left[H(\Tr(\rho E_b))-\sum_ip_iH(\Tr(\rho_i E_b))\right], 
\end{equation}
\normalsize
where where $H(\Tr(\rho E_b))$ denotes the Shannon entropy $-\sum p_b\log p_b$ of the probability vector with elements $p_b=\Tr(\rho E_b)$ [similarly for $H(\Tr(\rho_i E_b))$],  $\mathcal{E}\equiv\{p_i,\sigma_i\}$ denotes the signal ensemble, $\rho_i=\mathcal{N}(\sigma_i)$, $\rho=\sum_ip_i\rho_i$, and $\{E_b\}$ be the POVM on $\mathcal{H}_{out}$.

On the other-hand, the Holevo – Schumacher – Westmoreland theorem provides the maximum information carrying Capacity $\mathrm{C}_{PE}(\mathcal{N})$ of a channel (also called the Holevo Capacity $\mathrm{C}_{Hol}(\mathcal{N})$) when inputs are restricted to product states but entangled measurements are permitted. This is given by:
\small
\begin{equation}
\mathrm{C}_{Hol}(\mathcal{N})=\sup_{\mathcal{E}}\left(S\left[\mathcal{N}(\sigma)\right]-\sum_ip_iS\left[\mathcal{N}(\sigma_i)\right]\right), 
\end{equation}
\normalsize
where $\sigma=\sum_ip_i\sigma_i$, and $S[\chi]=-\tr(\chi\log\chi)$ denotes the von Neumann entropy of the density matrix $\chi$. Here (and throughout the paper) logarithm is considered with base $2$. 

Interestingly, there exist quantum channels with $\mathrm{C}_{Hol}>\mathrm{C}_{Shan}$ which establishes superiority of entangled decoding in classical information transfer through quantum channels \cite{Fuchs17,Schumacher01,King02,Hayashi05}. More strikingly, some channel may require non-orthogonal input signals to achieve the value of $\mathrm{C}_{Hol}$ \cite{Fuchs17,Schumacher01}. Further, a qubit channel may require more than two input states to achieve $\mathrm{C}_{Hol}$ \cite{King02,Hayashi05}. Later, the result of Hastings \cite{Hastings09} indicates that further increase in Capacity may also be achieved whenever entanglement is allowed both in input and output, leading to the concept of ultimate capacity which is defined as:  
\begin{equation}
\mathrm{C}_{ult}(\mathcal{N})=\mathrm{C}_{EE}(\mathcal{N})=\lim\limits_{n \to \infty} \frac{1}{n} \sup\limits_{\mathcal{E}, \mathcal{M}} I^{q}_{\mathcal{N}^{\otimes n}}(\mathcal{E};\mathcal{M}).
\end{equation} 
Here the supremum is taken over all possible signals ensemble $\mathcal{E}$ in $\mathcal{H}_{in}^{\otimes n}$, and all possible measurements $\mathcal{M}$ on $\mathcal{H}_{out}^{\otimes n}$. For a fixed ensemble $\mathcal{E}=\{p_i,\sigma_i\}$  and a POVM $\mathcal{M}=\{E_b\}$ we have $I^{q}_{\mathcal{N}}(\mathcal{E};\mathcal{M}):=S\left(\tr[\mathcal{N}(\sigma)E_b]\right)-\sum_ip_iS\left(\tr[\mathcal{N}(\sigma_i)E_b]\right)$.

In general, these capacities for a channel are expected to characterize the efficiency of channels in different scenarios. However, in this paper we show that for certain pairs of noisy quantum channels, none of these capacities can characterize an order of merit to delineate their utility in classical information tasks. Towards this aim, we consider a guessing game as studied in \cite{Frenkel15,Klobuchar10}. But before presenting our main result we first recall three qubit channels that will be relevant for our study. 

I. Depolarizing channel ($\mathcal{D}_p$): Given an arbitrary input state $\rho\in\mathrm{D}(\mathbb{C}^2)$, the channel maps it into the maximally mixed state with probability $p$ and keeps the state intact with the probability ($1-p$), {\it i.e.} $\mathcal{D}_p:~\rho\mapsto p~\mathbf{I}/2+(1-p)\rho$. Geometrically, this can be viewed as a shrinking of the Bloch sphere symmetrically from all directions. The Shannon capacity of this channel is given by \cite{Wilde13},
\begin{equation}
\mathrm{C}_{Shan}(\mathcal{D}_p)=1+\left(1-\frac{p}{2}\right)\log\left(1-\frac{p}{2}\right)+\frac{p}{2}\log\frac{p}{2}.
\end{equation}

For this channel, all the three classical capacities are same, {\it i.e.}  $\mathrm{C}_{Shan}(\mathcal{D}_p)=\mathrm{C}_{Hol}(\mathcal{D}_p)=\mathrm{C}_{ult}(\mathcal{D}_p)$ \cite{King03}. More generally Holevo Capacity in known to be additive for unital qubit channels \cite{King02(1)}. 

II. Splaying  channel ($\mathcal{S}$): The action of this channel on an arbitrary pure qubit state $\rho_{\alpha\beta}=1/2(\mathbf{I}+\vec{r}_{\alpha\beta}.\vec{\sigma})$, with $\vec{r}_{\alpha\beta}=(\cos\alpha \sin\beta , \sin \alpha \sin \beta , \cos \beta)$ is given by $\mathcal{S}(\rho_{\alpha\beta}):=1/2(\mathbf{I}+\vec{f}_{\alpha\beta}.\vec{\sigma})$, where $\vec{f}_{\alpha\beta}=1/3(1+\cos\alpha \sin\beta , \sqrt{3}\sin \alpha \sin \beta , 0)$ and $\vec{\sigma}:=(\sigma_x,\sigma_y,\sigma_z)$. This is an example of non-unital channel. 

The Shannon Capacity for splaying channel is given by $\mathrm{C}_{Shan}(\mathcal{S})=0.255992$ which is achieved for orthogonal input signals. Interestingly, for this channel $\mathrm{C}_{Hol}(\mathcal{S})$ is achieved with non orthogonal input states, and the vulue is given by $\mathrm{C}_{Hol}(\mathcal{S})=0.268932>\mathrm{C}_{Shan}(\mathcal{S})$ \cite{Fuchs17}.

III. King-Nathanson-Ruskai (KNR) Channel ($\mathcal{K}$): The action of this channel on a generic pure qubit state is given by, $\mathcal{K}(\rho_{\alpha\beta}):=1/2(\mathbf{I}+\vec{f}_{\alpha\beta}.\vec{\sigma})$, where $\vec{f}_{\alpha\beta}=\left(s\cos\alpha \sin\beta,s\sin\alpha \sin\beta,(1-\mu)+\mu\cos\beta \right)$, where $\mu\in[0,1]$ and $\mu\le s\le \sqrt{\mu}$. In this work, we will consider $\mu=0.5$ and $s=0.6$. For these particular choices of parameters, the Shannon Capacity is given by $\mathrm{C}_{Shan}(\mathcal{K})=0.321928$ (achieved with orthogonal inputs), whereas $\mathrm{C}_{Hol}(\mathcal{K})=0.32499>\mathrm{C}_{Shan}(\mathcal{K})$ achieved with three non orthogonal input signals \cite{King02}.  

%\section{Dollar Bill Capacity}
{\it Expected Payout.--} Here we first recall the guessing game introduced in \cite{Frenkel15,Klobuchar10}. The game $\mathcal{G}(N)$ involves two spatially separated players Alice and Bob, and a Referee who puts $\$1$ bill uniformly random into one of $N$ boxes. Bob can pick one of the boxes and can earn the contents of the box. Alice knows where the $\$1$ bill has been placed and she wishes to help Bob. However, only a limited classical or quantum communication is allowed from Alice to Bob. It has been shown that quantum communication is no better than classical communication in providing a higher expected value of the money won \cite{Frenkel15}. Importantly, this {\it no-go} result is not imposed by Holevo theorem and thus should be considered as a separate {\it no-go} statement limiting the classical information storage capacity of a quantum system. 

The authors in \cite{Frenkel15} have considered perfect quantum communication between Alice and Bob. Here we study the scenario where the quantum channel from Alice to Bob is noisy. For such a channel $\mathcal{N}$, Alice and Bob try to follow the best encoding-decoding scheme to obtain the highest expected value of the money won. Given that the $\$1$ bill is in the $i^{th}$ box, Alice follows an encoding $i\mapsto\rho_i$ and sends the state to Bob through the channel $\mathcal{N}$. Bob Performs some N-outcome POVM $M=\{E_i~|~E_i\ge0,~\&~\sum_{i=1}^NE_i=\mathbf{I}\}$ on the received state $\mathcal{N}(\rho_i)$ and upon obtaining outcome `$i$' he chooses the $i^{th}$ box. This motivates us to define a quality of merit, namely the expected payout, of a noisy channel in playing this particular game. For a given channel $\mathcal{N}$, the expected payout $\theta_{\$}(\mathcal{N})$ is defined as,
\begin{equation}
\theta_{\$}(\mathcal{N}):=\max_{\mathcal{E},M}\frac{1}{N}\sum_{i=1}^N\mbox{Tr}\left[E_i\mathcal{N}(\rho_i)\right].
\end{equation} 
While playing this $\$$-bill game, consideration of noisy quantum channel has also been studied in a recent work \cite{DallArno20}. There the authors reported optimum values of the above quantity and several other related quantities in closed form for several classes of quantum channels. In the following we will, however, show that the quantity $\theta_{\$}$ does not follow the ordering of the quantity $\mathrm{C}_{\pounds}$, where $\pounds\in\{Shan, Hol, ult\}$. 
\begin{center}

{\it When noisier is better.--} We consider the $\$$-bill game $\mathcal{G}(3)$ and calculate the expected payout $\theta_{\$}$ for the aforementioned three qubit channels (see the Table \ref{tab1}).
\begin{table}[b!]
\begin{tabular}{|c|c|c|c| } 
\hline
$\mathcal{N}$&$\theta_{\$}$& Encoding& Decoding\\
\hline\hline
\multirow{2}{2em}{$~\mathcal{D}_p$}& \multirow{2}{3em}{$~~\frac{2-p}{3}$} & $\rho_1=\rho_2=\frac{1}{2}(\mathbf{I}+\sigma_z),$&$\{E_1=\rho_1,E_2=0,$\\ 
&  & $\rho_3=\frac{1}{2}(\mathbf{I}-\sigma_z)$&$~~~~~~~~~~~~E_3=\rho_3\}$\\
\hline
\multirow{2}{2em}{$~\mathcal{S}$}& \multirow{2}{3em}{$\frac{3+\sqrt{3}}{9}$} & $\rho_1=\rho_2=\frac{1}{2}(\mathbf{I}+\sigma_y),$&$\{E_1=\rho_1,E_2=0,$\\ 
&  & $\rho_3=\frac{1}{2}(\mathbf{I}-\sigma_y)$&$~~~~~~~~~~~~E_3=\rho_3\}$\\
\hline
\multirow{2}{2em}{$~\mathcal{K}$}& \multirow{2}{3em}{$~~\frac{8}{15}$} & $\rho_1=\rho_2=\frac{1}{2}(\mathbf{I}+\sigma_x),$&$\{E_1=\rho_1,E_2=0,$\\ 
&  & $\rho_3=\frac{1}{2}(\mathbf{I}-\sigma_x)$&$~~~~~~~~~~~~E_3=\rho_3\}$\\
\hline
\end{tabular}
\caption{Expected Payout in $\mathcal{G}(3)$ for qubit channels $\mathcal{D}_p$, $\mathcal{S}$, and $\mathcal{K}$. One optimal encoding and decoding strategy for each channel is enlisted.}\label{tab1}
\end{table}
\end{center}
\begin{figure}[t!]
\centering
\includegraphics[height=6cm,width=9cm]{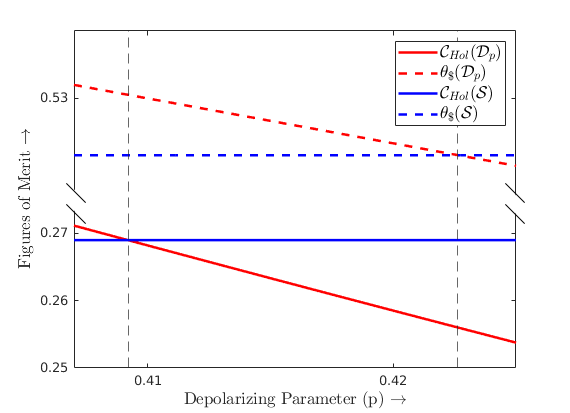}
\caption{For $0.4092<p<0.4226$ we have $\mathrm{C}_{Hol}(\mathcal{D}_p)<\mathrm{C}_{Hol}(\mathcal{S})$ but $\theta_{\$}(\mathcal{D}_p)>\theta_{\$}(\mathcal{S})$.}
\label{fig1}
\end{figure}
\begin{figure}[b!]
\centering
\includegraphics[height=6cm,width=9cm]{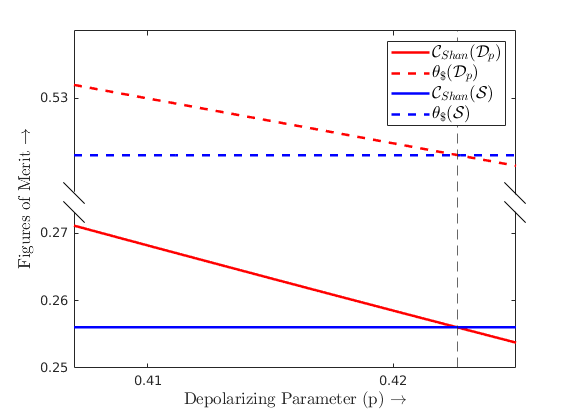}
\caption{There is no reversal in ordering. For all $p$  $\mathrm{C}_{Shan}(\mathcal{D}_p)<\mathrm{C}_{Shan}(\mathcal{S})\implies$  $\theta_{\$}(\mathcal{D}_p)<\theta_{\$}(\mathcal{K})$.}
\label{fig2}
\end{figure}

A straightforward comparison yields $\mathrm{C}_{Hol}(\mathcal{D}_p)<\mathrm{C}_{Hol}(\mathcal{S})$ whenever $p>0.409235$ (see Fig. \ref{fig1}). On the other hand, for $p<0.42265$ we have $\theta_{\$}(\mathcal{D}_p)>\theta_{\$}(\mathcal{S})$. Therefore, quite surprisingly, within the depolarising parameter range $p\in(0.409235,0.42265)$ the depolarising channel is preferable over the splaying channel for playing the $3$-box dollar game even though Holevo Capacity of the former channel is strictly lesser than the later. The same surprising feature appears even if we consider ultimate Capacity of these two channels instead of their Holevo Capacity. This is because $\mathrm{C}_{ult}(\mathcal{D}_p)=\mathrm{C}_{Hol}(\mathcal{D}_p)<\mathrm{C}_{Hol}(\mathcal{S})\le\mathrm{C}_{ult}(\mathcal{S})$. However, while considering their Shannon capacity this fact vanishes as $\mathrm{C}_{Shan}(\mathcal{D}_p)>\mathrm{C}_{Shan}(\mathcal{S})$ for $p<0.42265$ which is also the case for their expected payout (see Fig. \ref{fig2}). This gives an impression that the aforesaid reverse ordering may arises due to the fact that $\theta_{\$}$, likewise $\mathrm{C}_{Shan}$, characterizes utility of the channel in single shot regime whereas $\mathrm{C}_{Hol}$ and $\mathrm{C}_{ult}$ capture utility in multi-shot regimes. However, our next example shows that his is not the case. 

\begin{figure}[t!]
\centering
\includegraphics[height=6cm,width=9cm]{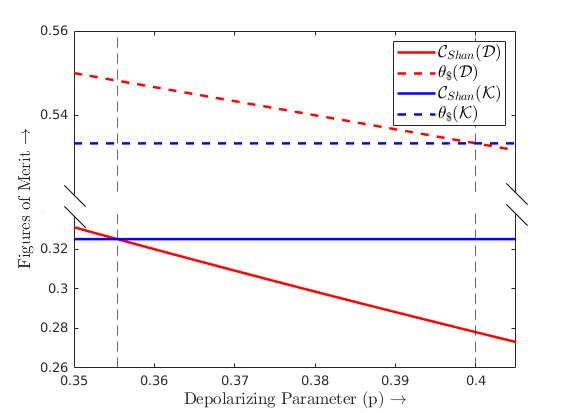}
\caption{For $0.358169<p<0.4$ we have $\mathrm{C}_{Shan}(\mathcal{D}_p)<\mathrm{C}_{Shan}(\mathcal{K})$ but $\theta_{\$}(\mathcal{D}_p)>\theta_{\$}(\mathcal{K})$.}
\label{fig3}
\end{figure}

We now consider the KNR channel. For the depolarizing parameter range $p\in(0.358169,0.4)$ we have $\theta_{\$}(\mathcal{D}_p)>\theta_{\$}(\mathcal{K})$ even though in this range $\mathrm{C}_{Shan}(\mathcal{D}_p)<\mathrm{C}_{Shan}(\mathcal{K})$ (see Fig. \ref{fig3}). It follows that this reverse ordering remains even when we consider the multi-shot capacities, as $\mathrm{C}_{ult}(\mathcal{D}_p)=\mathrm{C}_{Hol}(\mathcal{D}_p)=\mathrm{C}_{Shan}(\mathcal{D}_p)<\mathrm{C}_{Shan}(\mathcal{K})<\mathrm{C}_{Hol}(\mathcal{K})\le\mathrm{C}_{ult}(\mathcal{K})$. In fact, if we consider the Holevo capacity then the range of reversal widens to $p\in(0.355391,0.4)$. In the next we look for the origin of this peculiar behavior.  

%\section{Lossy Compression in Classical Channels}
{\it Lossy Compression in Classical Channels.--}
The $3$-box $\$$-bill game requires $\log_23$-bits of communication from Alice to Bob for perfect success. Since only $1$-bit communication is allowed, this is a lossy compression scenario. For such a scenario the {\it rate distortion theory} yields the relative efficiency of a channel \cite{Cover06}. 

Given a certain input distribution, the average distortion can be calculated using the Hamming distortion measure defined as
\begin{equation}
d_N(x_N,\hat{x}_N)= 
\begin{cases}  
0 & \text{if } x_N = \hat{x}_N, \\
1 & \text{if } x_N \neq \hat{x}_N; 
\end{cases}\\
\end{equation}
where $(x_N,\hat{x}_N)$ be the source alphabet-reproduction alphabet pairs with N being the block length of the code used. The distortion rate function is  defined as $\mathbb{D}_N(\mathrm{R}_N):=\inf~\Bar{d}_N(x_N,\hat{x}_N)$, where $\Bar{d}_N(x_N,\hat{x}_N)= E(d_N(x_N,\hat{x}_N))$ denotes the expectation value of the distortion (average distortion). Let us denote, $\Bar{d}:= \lim_{N \to \infty} \frac{1}{N} \Bar{d}_N$. 

Ergodic hypothesis allows us to interpret the the average distortion as $P_{loss}$, where $P_{loss}=\sum_{i=1}^{3} P\left(loss|\hat{i}~\right)P(\hat{i}~)$. Here, $\hat{i}$ is the Bob's guess of which box the dollar bill was initially placed. In the game we have considered, $P_{loss}$ is bounded below by 1/3 \cite{Frenkel15,Klobuchar10}. Therefore, $\Bar{d}(x,\hat{x})$ is minimized when all the input variables occur with equal probability implying $\mathbb{D}(\mathrm{R})=\Bar{d}(x,\hat{x})$. This holds as long as the channel used doesn't introduce any error that cannot be corrected with some error correcting method and is therefore independent of the channel used. 

Finally, we recall here that $\mathbb{D}(\mathrm{R})$ is a monotonic function decreasing in $\mathrm{R}$. Also, the rate is bounded above by the transmission rate of the channel -- the channel Capacity $\mathrm{C}$. Thus, two channels with Capacity $\mathrm{C}_1 > \mathrm{C}_2$ implies that the maximum rate through them are $\mathrm{R}_{1}^{\max}>\mathrm{R}_{2}^{\max}$, and consequently this implies $\mathbb{D}_1<\mathbb{D}_2$. Since $\mathbb{D}=P_{loss}$ for our case, therefore $\mathrm{C}_1>\mathrm{C}_2$ implies $P_{loss}^1>P_{loss}^2$; and hence classical channels with higher Capacity always perform better in the game in consideration.

%\section{Discussion}
{\it Discussion.--} Channel capacities are adequate characterizations of classical channels in deciding their order of merit for information theoretic tasks. In the lossless encoding-decoding scheme it is defined using the maximum transfer rate at which all errors due to the channel can be made arbitrarily small by suitable error correction schemes. Thus, higher transfer rate implies more transfer of useful information. Additionally, it is known to hold true for lossy compression schemes too. In the case of quantum channels, for lossless encoding-decoding, Classical Capacity of the channel again characterizes the order of merit of the channels in information theoretic tasks as in the case of classical channels. However, the observations made in the present paper demonstrate that this may not be the case for lossy compression schemes in quantum channels. 

We have shown that - for sufficiently large range of depolarizing parameter - while Shannon capacity of a qubit depolarizing channel is strictly less than that of the KNR Channel, this doesn't imply that the later channel is always preferable than the former for transferring classical information. In fact we have shown that in a two party guessing game the depolarizing channel, in the same depolarizing parameter range, is more advantageous than the KNR channel despite its lower Shannon capacity. This result is only further strengthened when we consider the Holevo capacity or the ultimate capacity. We do not expect that this is an isolated example. It remains an interesting open question whether this phenomenon persists when the expected payout is optimized over entangled encoding and decoding. Our result thus adds a new dimension in the study of quantum channels by making their classical utility context dependent. We therefore posit that the utility of channels can no longer be characterized by a single measure of Capacity. 

\begin{acknowledgments} 
We would like to thank Mohammad Alhejji for his useful comments on the earlier version of the manuscript. MB thankfully acknowledges discussions with Mihály Weiner (through email) regarding the $\$$-Bill game. RT acknowledges support through The Institute for Quantum Information and Matter which is an NSF Physics Frontiers Center. GS was partially supported by NSF CAREER award CCF 1652560. MB acknowledges support through the research grant of INSPIRE Faculty fellowship from the Department of Science and Technology, Government of India.  
\end{acknowledgments}

\end{document}